\documentclass[twocolumn,showpacs,preprintnumbers,amsmath,amssymb]{revtex4}

\usepackage{graphicx}
\usepackage{amsmath}
\usepackage{bm}
\usepackage{multirow}
\usepackage{comment}

\newcommand{\simgt}{\lower.5ex\hbox{$\; \buildrel > \over \sim \;$}}
\newcommand{\simlt}{\lower.5ex\hbox{$\; \buildrel < \over \sim \;$}}
\newcommand{\largesmall}{\lower.5ex\hbox{$\; \buildrel {\LARGE >} \over {\small 
<}$}}

\begin{document}
\title{
Breaking the scale invariance of the primordial power spectrum in \\Ho\v{r}ava-Lifshitz cosmology}


\author{
Kazuhiro {Yamamoto}$^{1}$, Tsutomu Kobayashi$^{2}$, Gen Nakamura$^{1}$}

\affiliation{
$^{1}$Department of Physical Science, Hiroshima University,
Higashi-Hiroshima 739-8526,~Japan
\\
$^{2}$Department of Physics, Waseda University,
Okubo 3-4-1, Shinjuku, Tokyo, 169-8555,~Japan
}



\begin{abstract}
We study the spectral tilt of primordial perturbations
in Ho\v{r}ava-Lifshitz cosmology.
The uniform approximation, which is a generalization of the familiar
Wentzel-Kramers-Brillouin (WKB) method, is employed to compute the spectral index
both numerically and analytically in a closed form.
We clarify how the spectral index depends on the inflation model and
parameters in the modified dispersion relation.
\end{abstract}

\preprint{HUPD-0903, WU-AP/301/09}
\maketitle

\section{Introduction}

A power-counting renormalizable theory of gravity,
proposed recently by Ho\v{r}ava~\cite{Horava}, has attracted much attention
and generated a trend in quantum gravity.
The essential aspect of the theory is broken Lorentz invariance
in the ultraviolet (UV), where it exhibits a Lifshitz-like anisotropic scaling,
$t\to \ell^z t$, $\Vec{x}\to\ell\Vec{x}$, with the dynamical critical exponent $z=3$.
Modification to general relativity thus introduced in the high energy regime
brings interesting consequences in cosmology~\cite{Calcagni},
including the generation of chiral gravitational waves from inflation~\cite{TakahashiSoda},
a bouncing scenario~\cite{Brandenberger},
a possible candidate for dark matter~\cite{Darkmatter}, and others~\cite{MNTY}.
Since signals from the early universe
could be observed for instance through the cosmic microwave background (CMB)
temperature anisotropies and the stochastic gravitational wave background,
it is important to study the dynamics cosmological perturbations in 
Ho\v{r}ava-Lifshitz (HL) cosmology (see, e.g., Ref.~\cite{WangMaartens} and 
references therein). However, it should be noted that problems of
HL gravity are still under debate~\cite{Darkmatter,Charmousis,
BogdanosSaridakis}.

One of the basic ingredients in today's cosmology is
the {\em almost} scale-invariant power spectrum of primordial perturbations
generated quantum mechanically in the early universe,
and what is equally important is {\em a small
deviation from the exact scale-invariance.}
Inflation involves a natural mechanism to generate almost
scale-invariant cosmological perturbations,
and indeed this fact together with observations of the CMB and
the large scale structure (LSS) in the universe provides increasing support
for inflation.
The scale-invariance in this case
essentially relies upon the almost constant Hubble rate,
and a slight deviation from the exact de Sitter expansion
is responsible for the small tilt of the spectrum.

Another interesting mechanism
to generate scale-invariant cosmological perturbations
was recently pointed out by Mukohyama~\cite{Mukohyama}
in connection with HL cosmology.
(See also Refs.~\cite{CP1,CP2,CP3,Gao,CaiZhang, KUY} for cosmological perturbations
in HL gravity.)
The anisotropic scaling in the UV allows us to include
higher spatial derivative terms in a scalar field Lagrangian,
modifying the dispersion relation into the form $\omega^2\propto k^6$
in the early universe,
which results in
scale-invariant scalar field perturbations.
In contrast to the standard mechanism,
the spectrum in Mukohyama's scenario
does not depend on the Hubble rate explicitly but is determined only through
some model parameter.
Therefore, the new mechanism liberates inflation models from those with
a very flat potential.
In Mukohyama's scenario, the dispersion relation
which is not exactly of the form $\omega^2\propto k^6$ will break
the scale-invariance.

In this paper we compute a slightly tilted spectrum of quantum fluctuations
generated in HL cosmology,
evaluating accurately the spectral index by
means of the uniform approximation~\cite{Martin,HabibA,HabibB}.
The uniform approximation is a generalized  
Wentzel-Kramers-Brillouin (WKB) method, which is 
mathematically controlled and systematically extendable.
To the best of the authors' knowledge,
the uniform approximation has not been used so far for
models with modified dispersion relations.
We consider a power-law/slow-roll inflationary background in HL cosmology,
which provides a transparent example that allows for
analytical and numerical calculations.
Although our primary motivation comes from HL cosmology,
we believe that the approach presented in this paper
has a wide range of applications in different scenarios of cosmology.



\section{Uniform approximation}

We consider a quantum field whose Fourier coefficients obey
the equation of motion,
\begin{eqnarray}
\varphi_k''(\eta)
+\left[k^2_{\rm eff}(\eta)-{a''\over a}\right]\varphi_k(\eta)=0,
\label{starteq}
\end{eqnarray}
where the prime denotes the differentiation with respect to conformal time $\eta$,
$ a(\eta)$ is the scale factor,
and $k^2_{\rm eff}(\eta)$ is the {\it effective}
frequency~\cite{TakahashiSoda,Mukohyama}.
The field $\varphi_k$ will be a gravitational wave
(multiplied by the scale factor)~\cite{TakahashiSoda}
and a scalar field perturbation (multiplied by a scale factor)~\cite{Mukohyama}.
Following the prescription
of the uniform approximation~\cite{Martin,HabibA,HabibB}, 
we rewrite the above equation as
\begin{eqnarray}
\varphi_k''(\eta)
=\left[g(\eta)+q(\eta)\right]\varphi_k(\eta),
\end{eqnarray} 
where we defined 
\begin{eqnarray}
&&g(\eta)=-k^2_{\rm eff}(\eta)+{a''\over a}+{1\over 4\eta^2},
\end{eqnarray}
and $q(\eta)=-1/4\eta^2$, which 
guarantees the convergence
of the approximation. 
This method provides a single approximate solution 
$  \varphi_k(\eta)=[g(\eta)/\zeta(\eta)]^{-1/4}
  \left[\alpha_k {\rm Ai}(\zeta)+\beta_k{\rm Bi}(\zeta)\right]$,
with
$\zeta(\eta):=
\pm\left[{\pm(3/2)}\int_{\bar\eta}^\eta\sqrt{\pm g(\eta_1)}
d\eta_1\right]^{2/3}$,
where $+$ for $\eta >\bar\eta$, $-$ for $\eta <\bar\eta$, respectively, 
and $\bar\eta$, which corresponds to the {\it effective} 
horizon-crossing time, is defined by
$g(\bar\eta)=0$,
and ${\rm Ai}(z)$ and ${\rm Bi}(z)$ are the Airy functions.
Choosing
$\alpha_k= \sqrt{\pi / 2}$ and $\beta_k= -i\sqrt{\pi / 2}$,
the mode function $\varphi_k$ is made to satisfy the Wronskian normalization condition
$\varphi_k\varphi^*_k{}'-\varphi_k{}'\varphi^*_k=i$
and has the asymptotic early-time behavior
$\varphi_k(\eta)\simeq[2k_{\rm eff}(\eta)]^{-1/2}\times
\exp[-i\int^\eta_{\bar\eta} k_{\rm eff}(\eta_1)d\eta_1]$,
for $\eta\ll \bar\eta$.

Using the asymptotic formula of the Airy function
${\rm Ai}(z)\simeq (2\sqrt{\pi})^{-1}z^{-1/4}e^{-2z^{3/2}/3}$
and
${\rm Bi}(z)\simeq (\sqrt{\pi})^{-1}z^{-1/4}e^{2z^{3/2}/3}$
for $|z|\rightarrow \infty$ with $-\pi<{\rm arg}z<\pi$, we
find the asymptotic formula of
the mode function after the horizon crossing:
\begin{eqnarray}
\varphi_k(\eta)\simeq-{i\over \sqrt{2}} [g(\eta)]^{-1/4} \exp\left[
\int_{\bar\eta}^\eta \sqrt{g(\eta_1)}d\eta_1
\right].
\end{eqnarray}
Thus, we finally obtain the power spectrum $P_\varphi(k,\eta)
={\left|{\varphi_k(\eta)/ a(\eta)}\right|^2}$, 
and the spectral index of the quantum fluctuations
can be estimated from $d\ln P_\varphi(k,\eta)/d\ln k$,
which leads to
\begin{eqnarray}
 n_{\rm ini}
&\simeq&
 \lim_{\eta\rightarrow\eta_{\rm max}} 2k{d\over dk} \int_{\bar\eta(k)}^\eta 
 \sqrt{g(\eta_1)}d\eta_1.
\label{sectralindex}
\end{eqnarray} 
This provides a closed formula for the spectral index. Note, however,
that this formula cannot be used for the cases in which the fluctuations 
reenter the horizon and the horizon crossing takes place again.

\def\betaH{{H_*}}
\section{Power-law inflation}
\begin{figure}[tb]
\begin{center}
\includegraphics[width = 80mm,keepaspectratio,clip]{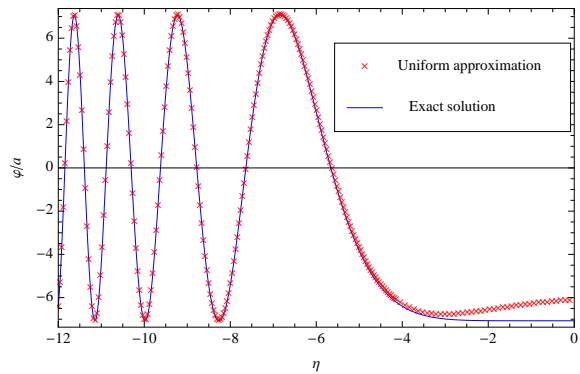}
\caption{Growing mode solution to Eq.~(\ref{starteq}), obtained by
employing the uniform approximation, is compared to the exact result.
The parameters are given by $\mu=2$, $n=3$, $H_*=1$, and $k^3/M^2=10^{-2}$.
The horizon-crossing time is $\bar\eta = -4.14$.}
\label{fig-compare}
\end{center}
\end{figure}

For simplicity, 
let us consider power-law inflation~\cite{LucchinMa},
for which
$ a(t)\propto t^{1+n}$ in terms of proper time $t$ and 
$ a(\eta)=1/ (-\betaH\eta)^{1+1/n}$ in terms of conformal time 
$\eta ~(-\infty<\eta<0)$, 
with $ n>0$. We first study a simple case, 
\begin{eqnarray}
k^2_{\rm eff}(\eta)=k^2\left(
{k^2\over M^2 a^2}\right)^\mu. 
\label{keffpowerlow}
\end{eqnarray}
In this case, it is easy to find
the superhorizon behavior in the uniform approximation:
\begin{eqnarray}
&&\varphi_k(\eta)\approx -i a(\eta){\cal A}e^{-\nu}(3+2/n)^{\nu-1/2}k^{-(1+\mu)\nu },
\label{amplitudea}
\\
&&
{\cal A}:=
H^{\nu-1/2}_*
M^{\mu\nu},
\;\;\nu:=\frac{1+3n/2}{\mu+n(\mu+1)},\nonumber
\end{eqnarray}
where ${\cal A}$ defines the typical amplitude of the fluctuation.
In the de Sitter limit $n\to\infty$ with the usual dispersion relation $\mu=0$,
we have ${\cal A}=H_*$.
For $\mu=2$ we find ${\cal A}=M$,
and thus the result of~\cite{Mukohyama} is reproduced.
Fortunately, for the dispersion relation~(\ref{keffpowerlow}),
Eq.~(\ref{starteq}) can be solved exactly,
and so we can compare the exact result with the approximate one.
The appropriately normalized solution to Eq.~(\ref{starteq}) is given by
\begin{eqnarray}
\varphi_{{\rm exact}}(\eta) = \frac{1}{2}\sqrt{\frac{n\pi}{\mu+n(\mu+1)}}
(-\eta)^{1/2} H^{(1)}_{\nu}(y(\eta)),
\end{eqnarray}
where $H_\nu^{(1)}$ is the Hankel function of the first kind and
$y(\eta) := \int^{0}_\eta k_{{\rm eff}}(\eta_1) d\eta_1$.
This then implies that $|\varphi_k |= {\cal C}|\varphi_{{\rm exact}}|$
at $\eta\approx 0$, where
$
{\cal C}:=\sqrt{2\pi} {e^{-\nu}\nu^{\nu-1/2}}/{\Gamma(\nu)}.
$
In the de Sitter limit, $n\to\infty$, for $\mu=0$ 
we find ${\cal C} = 3\sqrt{2}e^{-3/2}\simeq 0.95$,
while for $\mu=2$ we have ${\cal C}=\sqrt{2/e}\simeq 0.86$ irrespective of $n$.
It is important to note that ${\cal C}$ gives rise to a $k$-independent
correction to the amplitude, and therefore this does not affect the
computation of the spectral index.
The exact result and the approximate one are compared in
Fig.~\ref{fig-compare} in the case of $\mu=2$.

Now we are in position to study the spectral index for
the above model. Eq.~(\ref{sectralindex}) or (\ref{amplitudea}) gives
\begin{eqnarray}
 n_{\rm ini}+3=
\frac{\mu-2}{\mu+n(\mu+1)},
\label{nspecform}
\end{eqnarray}
which leads to the scale-invariant spectrum 
$ n_{\rm ini}=-3$ when $\mu=2$, as expected~\cite{Mukohyama}.
{The result~(\ref{nspecform}) is consistent with
what was obtained in a different way by matching two asymptotic solutions
at horizon crossing~\cite{CaiZhang}.}
Since the slow-roll parameter $\epsilon\;(:=-H'/aH^2)$ is explicitly given by
$\epsilon = 1/(1+n)$ for power-law inflation, Eq.~(\ref{nspecform}) 
can be written as $ n_{\rm ini}+3\simeq -\epsilon({2-\mu})/({1+\mu})$
to leading order in the slow-roll parameter.
One can confirm that this reproduces the standard result for $\mu=0$.
Note also that $n_{\rm ini}$ approaches $-3$ in the limit $n\to\infty$ ($\epsilon\to 0$)
irrespective of the value $\mu$. 
This means that in order to break the scale invariance of the primordial spectrum 
we need the background universe which is not exactly de Sitter
and $\mu\neq 2$. 

Next, let us study the case of HL cosmology, for which
\begin{eqnarray}
k^2_{\rm eff}(\eta)=k^2\left(
{k^4\over M^4 a^4}+\alpha_1{k^3\over M^3 a^3}+\alpha_2{k^2\over M^2 a^2}+1
\right).
\label{defkeff}
\end{eqnarray}
In the case of gravitational waves, $\alpha_1$ is non-zero
\cite{TakahashiSoda}, while $\alpha_1=0$ for a scalar field \cite{Mukohyama}.
For this dispersion relation Eq.~(\ref{sectralindex}) reduces to
\begin{eqnarray}
  n_{\rm ini}&=&{\kappa n\over 1+n}
  \times\nonumber\\&&\hspace{-5mm}
  \int_0^{\bar z} 
  \frac{-z^{-(2+n)/(1+n)} [z^2f(z)]'dz }{
\sqrt{
\left(n^{-1}+3/2\right)^2
 \kappa^{-2} z^{-2n/(1+n)}
-f(z)}
},\label{integrationex}
\end{eqnarray}
where $f(z):=z^4+\alpha_1 z^3+\alpha_2z^2+1$,
$ \kappa:=(M/H_*)(k/M)^{1/(1+n)}$ and
we defined a new variable $z$ by $-\betaH \eta= (Mz/k)^{n/(1+n)}$.
The prime in Eq.~(\ref{integrationex}) stands for the differentiation with respect to $ z$.
The upper limit of integration $\bar z$ satisfies
\begin{eqnarray}
\left({1/n}+{3/2}\right)^2
 \kappa^{-2} \bar z^{-2n/(1+n)}-f(\bar z)=0.
\label{solbarz}
\end{eqnarray}

{ Figure~\ref{fig:contour2}
shows the contour of $n_{\rm ini}$ in the $\kappa$ and $1/n$ plane, 
obtained from a numerical integration of Eq.~(\ref{integrationex}). 
The parameters are given by $\alpha_1=0$ and $\alpha_2=1$ for the left panel, which
corresponds to a scalar field perturbation of~\cite{Mukohyama},
$\alpha_1=2, ~\alpha_2=1$
and $\alpha_1=-2, ~\alpha_2=1$, respectively, for the center and right panels.
The latter two cases correspond to chiral gravitational waves
analyzed in~\cite{TakahashiSoda}.

The result is understood in an analytic way as follows. 
One can recast the expression~(\ref{defkeff}) in
$k_{\rm eff}^2\propto k^{2+2\mu(\eta, k)}$, 
where
$\mu(\eta, k)=(1/2){d\ln f}/{d\ln z}$,
and so $\mu$ is dependent on $\eta$ and $k$ only through $z$.
For the case $\kappa \ll 1$, the solution to Eq.~(\ref{solbarz}) is such that 
$\bar z\gg 1$, and thus we have
$\bar z\simeq [\kappa^{-1}(n^{-1}+3/2)]^{(1+n)/(2+3n)}$
and 
$\mu(k,\bar\eta)\simeq 2-\alpha_1/(2\bar z)+(\alpha_1^2-2\alpha_2)/(2\bar z^2)$.
Then, Eq.~(\ref{nspecform}) gives 
\begin{eqnarray}
  n_{\rm ini}+3
\simeq -\left({\alpha_1\over 2\bar z}
-{\alpha_1^2-2\alpha_2\over 2\bar z^2}\right){1\over 2+3n}. 
\end{eqnarray}
This result applies to the case where the horizon crossing 
occurs in the UV regime, $k_{\rm eff}^2\sim k^6$. 
While,
for the case $\kappa \gg1$, the 
solution to Eq.~(\ref{solbarz}) is given by 
$\bar z\ll 1$, 
which corresponds to the case where the horizon crossing 
takes place in the infrared (IR) regime, $k_{\rm eff}^2\sim k^2$. 
In this case, we may write
$\bar z\simeq[\kappa^{-1}(n^{-1}+3/2)]^{(1+n)/n}$
and 
$
\mu(k,\bar \eta)\simeq \alpha_2 \bar z^2 +(3/2)\alpha_1 \bar z^3.
$
Then, Eq.~(\ref{nspecform}) yields
\begin{eqnarray}
  n_{\rm ini}\simeq -3\left(1-
{\alpha_2 \bar z^2 +3\alpha_1 \bar z^3/2\over n}
\right)\left(1+{2\over 3n}\right).
\end{eqnarray}
The approximate formula reproduces the qualitative behavior 
of the numerical result.
The above results suggest that 
we must require small $\kappa\;(\lesssim 1)$ or large $n\;(\gtrsim 10)$,
in order to have
{a spectrum that is almost scale-invariant but slightly
deviated from the scale-invariant one.}
Note that
\begin{eqnarray}
  \kappa&=&{10^{6}\over (1.5\times 10^{57})^{1/(1+n)}} 
\left(M\over 10^{19} {\rm GeV}\right)^{n/(1+n)}\times
\nonumber\\&&\quad
\left(10^{15} {\rm GeV}\over H_*\right)
\left({k\over{\rm Mpc}^{-1} }\right)^{1/(1+n)}.
\end{eqnarray}

The sign of $\alpha_1$ signals different chiral modes of
primordial gravitational waves~\cite{TakahashiSoda},
and hence the spectral tilt of the gravitational waves
is different for right-handed and left-handed modes.
The above result shows explicitly
the dependence of the spectral index on model parameters.
Our result also gives a hint on the scenario of~\cite{Mukohyama}, 
in which quantum fluctuations of a scalar field can be the origin of
curvature perturbations
by promoting the former to the latter, e.g., via the curvaton mechanism~\cite{curvaton}.

\begin{figure*}
  \leavevmode
  \begin{center}
    \begin{tabular}{ c }
      \includegraphics[width=160mm,angle=0]{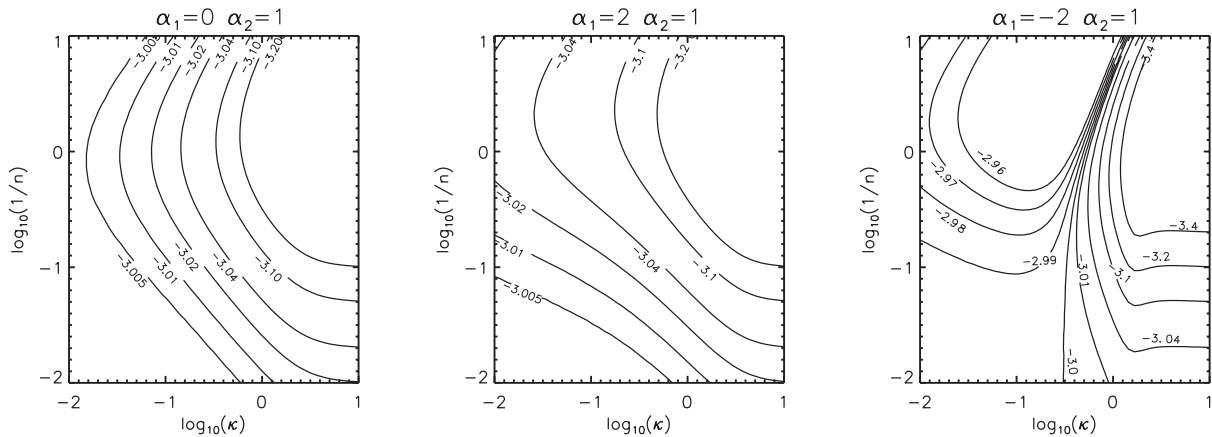}
    \end{tabular}
\caption{(a:~Left) Contours of $n_{\rm ini}$ on the $1/n$ and $\kappa$ plane. 
The parameters are $\alpha_1=0$ and $\alpha_2=1$.
The level of the contour is shown in the panel. 
(b:~Center) Same as (a), but with $\alpha_1=2$ and $\alpha_2=1$.
(c:~Right)
Same as (a), but with $\alpha_1=-2$ and $\alpha_2=1$.
\label{fig:contour2}
}
\end{center}
\end{figure*}

\section{Slow-roll inflation}

Let us move on to more general slow-roll inflation models.
We continue to study Eq.~(\ref{starteq}), but
now with $a''/a$ replaced by $a''/a\to
{C(\eta)^2}/{\eta^2}=2a^2H^2\left(1+E\right)$.
Here, $E$ is some general function whose explicit form depends 
on the problem under consideration:
$E=-\epsilon/2$ for a tensor perturbations and a scalar field perturbation
in the slow-roll regime;
$E\simeq\epsilon-3\delta/2$, where $\delta \;(:=-\ddot\phi/H\dot\phi)$
is another slow-roll parameter, for a scalar perturbation 
in conventional slow-roll inflation models, where 
$\phi$ is the inflaton field and the dot denotes the differentiation
with respect to the proper time.
However, a further careful investigation is necessary 
for a scalar (curvature) perturbation in HL cosmology.
For example, the $k^6$ term in the dispersion relation disappears
under the assumption of the detailed balance condition, 
according to the prescription of Ref.~\cite{Gao}.
References~\cite{WangMaartens, KUY} investigate
a different case
without the detailed balance condition but with the projectability condition.

Assuming the dispersion relation~(\ref{keffpowerlow}), a similar calculation 
for the right hand side of Eq.~(\ref{sectralindex}) yields
$ n_{\rm ini} +3\simeq -({3\epsilon})/({1+\mu})+\epsilon/3-4E/3$
to lowest order in the slow-roll parameters, where we assumed that $E$ is of 
order of the slow-roll parameters and used the relation 
$\eta\simeq 1/[-Ha(1-\epsilon)]$. 
In the case of $\mu=0$, the well-known result is reproduced
for the tensor and scalar perturbations.
{Similarly to the above, we can also derive formulas for the
dispersion relation~(\ref{defkeff}).}
Hereafter, we define $\bar z$ by the root of $f(\bar z)
=(\bar H^2/\bar z^2M^2)(9/4+2E-\epsilon/2)$, where $\bar{H}$ is 
the value of $H$ at the horizon-crossing time. 
For the case $M/\bar{H} \ll 1$, we find $\bar{z} \simeq (3\bar H/2M)^{1/3} \gg 1$ and
\begin{eqnarray}
 n_{\rm ini}+3 \simeq
  -\left(1+\frac{\alpha_1}{6\bar{z}}
  +\frac{2\alpha_2-\alpha_1^2}{6\bar{z}^2}\right)\epsilon
  +\frac{1}{3}\epsilon-\frac{4}{3}E,
\label{slowzgg}
\end{eqnarray}
while
for the case $M/\bar{H} \gg 1$ we have
$\bar{z} \simeq (3\bar H/2M) \ll 1$ and
\begin{eqnarray}
 n_{\rm ini}+3 \simeq
  -3\left(1-\alpha_2\bar{z}^2-\frac{3}{2}\alpha_1\bar{z}^3\right)\epsilon
  +\frac{1}{3}\epsilon-\frac{4}{3}E.
\label{slowzll}
\end{eqnarray}
One can check that these results are consistent with those of 
the power-law inflationary background with large value of $n+1\;(=1/\epsilon)$, 
by taking the expression for the gravitational waves and the scalar field, i.e., $E=-\epsilon/2$.

In~\cite{TakahashiSoda}, it was pointed out that
the evolution of
primordial gravitational waves is different depending on the parity mode.  
{}From Eqs.~(\ref{slowzgg}) and~(\ref{slowzll}), the difference of the spectral index is given by
\begin{eqnarray}
 \Delta n_{\rm ini}=
\left\{
\begin{array}{ll}
  -\displaystyle
{{\frac{\epsilon|\alpha_1|}{3}}
   {\left(\frac{2M}{3\bar{H}}\right)^{1/3}}}\; &(M/\bar{H} \ll 1)   \\
  9\epsilon|\alpha_1|\displaystyle
{\left(\frac{3\bar{H}}{2M}\right)^3}\; &(M/\bar{H} \gg 1) 
\end{array}
\right. .
\end{eqnarray}
The difference might not be negligible for $M/\bar{H} \ll 1$, 
which corresponds to the case where the horizon crossing takes place in the UV regime ($k_{{\rm eff}}^2\sim k^6$). 
However, the difference is small for $M/\bar{H} \gg 1$,
which is the case where
the horizon-crossing occurs in the IR regime ($k_{{\rm eff}}^2\sim k^2$). 
If we assume a scenario of Ho\v{r}ava-Lifshitz cosmology, 
one can predict the signal of gravitational waves, to be compared 
with observations (e.g., Ref.~\cite{Koh}).

Now we consider {curvature} perturbations in the
scenario of~\cite{Mukohyama}
by setting $\alpha_1=0$ and $E=-\epsilon/2$.
{Provided that the spectral tilt of scalar field perturbations is
directly translated into that of curvature perturbations,}
we have
\begin{eqnarray}
 n_{\rm ini}+3 \simeq
\left\{
\begin{array}{ll}
  \displaystyle
{-{\epsilon\alpha_2\over 3}{\left({2M\over 3\bar H}\right)^{2/3}}} \;&(M/\bar{H} \ll 1)   \\
  -2\epsilon+\displaystyle
{{27\epsilon\alpha_2\over 4}{\bar H^2\over M^2}} \;&(M/\bar{H} \gg 1) 
\end{array}
\right. .
\label{finalresult}
\end{eqnarray}
The CMB measurement by the WMAP satellite and LSS of 
galaxies provide a stringent constraint on the primordial spectral index. 
For example, the WMAP team reported 
$n_{\rm ini}+3\sim-0.04\pm0.013$ 
based on the $\Lambda$CDM model \cite{WMAP}, 
combined with Ia supernovae observations.  
This gives a constraint on the inflation model and the parameters in 
the scalar field Lagrangian, assuming that the almost 
scale-invariant spectrum we observe is originated from the 
Ho\v{r}ava-Lifshitz scalar field~\cite{Mukohyama}.

\section{Conclusions}
In this paper, we have studied the spectral tilt of 
quantum fluctuations in HL cosmology, employing 
the uniform approximation.
In the power-law inflationary background we have obtained the 
spectral index numerically and analytically. 
The deviation from the scale invariant spectrum is 
described by the parameters $\alpha_1$ and $\alpha_2$
of the model, as well as the background evolution, i.e., $n$. 
The case of a general slow-roll inflationary background was also investigated.
In order for Mukohyama's scenario~\cite{Mukohyama} to be successful,
the spectral index, given in terms of the model parameters by Eq.~(\ref{finalresult}),
is required to be $\sim -0.04$.

\acknowledgements
 
This work was supported by a Grant-in-Aid
for Scientific research of the Japanese Ministry of Education, 
Culture, Sports, Science and Technology (No.~21540270), 
and in part by the Japan Society for Promotion
of Science (JSPS) Core-to-Core Program ``International Research 
Network for Dark Energy.''
We would like to thank S. Mukohyama, G. Niz, T. Padilla,
and E. Saridakis for useful communications. 
TK is supported by the JSPS under Contract No.~01642.
He is grateful for
the kind hospitality of the theoretical astrophysics group at
Hiroshima University, where this work was initiated.

\end{document}